\documentclass[prl,reprint,twocolumn,showpacs]{revtex4}
\usepackage{amsmath,amssymb,graphicx,xcolor,units,hyperref}

\newcommand{\dd}[1]{\mathrm{d}#1\,}
\newcommand{\avg}[1]{\langle{#1}\rangle}
\renewcommand{\Re}{\mathop{\mathrm{Re}}}

\newcommand{\Tr}{\mathop{\mathrm{Tr}}}
\newcommand{\sgn}{\mathop{\mathrm{sgn}}}

\newcommand{\bra}[1]{\langle{#1}\rvert}
\newcommand{\ket}[1]{\lvert{#1}\rangle}

\begin{document}

\title{Microwave spectroscopy of Josephson junctions in topological superconductors}
\author{Pauli Virtanen}
\affiliation{O.V. Lounasmaa Laboratory, Aalto University,
  P.O. Box 15100,FI-00076 AALTO, Finland}
\affiliation{Institute for Mathematical Physics, TU Braunschweig,
  38106 Braunschweig, Germany}
\author{Patrik Recher}
\affiliation{Institute for Mathematical Physics, TU Braunschweig,
  38106 Braunschweig, Germany}
\date{\today}

\pacs{74.45.+c, 71.10.Pm, 73.23.-b}

\begin{abstract}
  We consider microwave spectroscopy of Josephon junctions composed of
  hybridized Majorana states in topological 1-D superconductors.
  We point out how spectroscopic features of the junction appear in
  the current phase relation under microwave irradiation.  Moreover,
  we discuss a way to directly probe the nonequilibrium state
  associated with the $4\pi$ periodic Josephson effect.
  In particular, we show how the microwave driving can be used to switch 
  from a $4\pi$ to a $2\pi$ Shapiro step in the current voltage relation.
\end{abstract}

\maketitle

Josephson junctions in topological superconductors differ from
conventional superconductor junctions in several fundamental ways.
\cite{kitaev2001-umf,kwon2004-faj,fu2009-jca,lutchyn2010-mfa} Their
low-energy Andreev bound state spectrum can generically be
well-separated from the continuum spectrum above the superconducting
gap. Moreover, the fermion parity structure of the states is
different, which is reflected e.g. in wave function overlaps. These
differences are due to the bound states being formed through
hybridization of Majorana states.

Several distinguishing features of these systems have been proposed,
\cite{fu2009-jca,kitaev2001-umf,dominguez2012-ddm,jiang2011-ujs,san-jose2012-aje,pekker2013-pcc,san-jose2013-mar}
and first experimental evidence of the relevant physics was recently
obtained in semiconductor
nanowires. \cite{mourik2012-smf,rokhinson2012-ofa} One major generic
feature is the $4\pi$ periodic ac Josephson
effect. \cite{kitaev2001-umf} It is a nonequilibrium effect, in which
the system retains memory of the population of the bound states during
time evolution of the superconducting phase difference
$\varphi\mapsto\varphi+2\pi$. This results to an effective $4\pi$
periodic current-phase relation, $I(\varphi)\sim{}I_c\sin(\varphi/2)$,
\cite{kitaev2001-umf} the consequences of which are visible in Shapiro
steps and other
observables. \cite{kitaev2001-umf,fu2009-jca,dominguez2012-ddm,jiang2011-ujs,san-jose2012-aje}
Importantly, the expected double-frequency Shapiro step feature was
recently seen in an experiment \cite{rokhinson2012-ofa}.

A well-established way to probe the spectrum of a quantum system
(e.g. a qubit) is to drive it, and look for resonances as a function
of the frequency of the drive. Driving stimulates transitions between
energy levels and thereby also induces a nonequilibrium state in the
system. This physics is in play in the Andreev bound states in
Josephson junctions
\cite{zazunov2003-alq,bergeret2010-tom,zgirski2011-elq} and Majorana
wires. \cite{schmidt2012-mqr,trif2012-rtm} Information obtained in
this way can also be useful in characterizing the special features of
topological superconductor junctions.  For instance, a $4\pi$ periodic
Josephson effect is not necessarily of a topological origin,
\cite{sau2012-pfa} even though it is a strong indication of it, but
combined with additional knowledge of the spectrum its accidental
occurrence can be excluded.

Here, we suggest how microwave driving can be used as a separate
control parameter for tuning the magnitude of the $4\pi$ periodic
Josephson current, which does not require crossing topological
transitions and thereby adjusting the level structure of the system,
which may complicate interpretation of the results. We also discuss
how the level structure of the junction is spectroscopically reflected
in the dc current in phase- and current-biased situations.

\begin{figure}
  \includegraphics{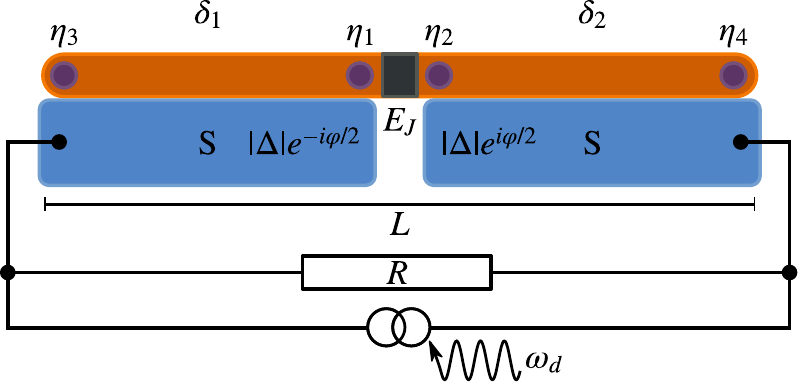}
  \caption{
    \label{fig:setup}
    (Color online) Josephson junction in a 1D (proximity-induced)
    topological superconductor.  The two Majorana states $\eta_1$,
    $\eta_2$ located at each side of the junction are hybridized
    (Josephson energy $E_J$), whereas the remaining two $\eta_3$,
    $\eta_4$ are spatially separated and weakly coupled (coupling
    energies $\delta_j$).  The system is assumed to be current biased and
    shunted with resistance $R$ (as shown) or phase biased ($\varphi$
    fixed), and driven with a finite frequency $\omega_d$.
  }
\end{figure}

\emph{Model.}  We describe the physics of the Josephson junction in
topological superconductor (TJJ) by a Bogoliubov--de-Gennes
Hamiltonian ${\cal H}_{BdG}(t)$.  The time-dependence originates from
the time dependence of the superconducting phase $\phi(x,t)$; we
assume (see Fig.~\ref{fig:setup}) that to the left of the junction
$\phi(x,t)=-\varphi(t)/2$, $x<0$, and to the right,
$\phi(x,t)=\varphi(t)/2$, $x>0$.  Such a Hamiltonian can be
conveniently rewritten in the corresponding instantaneous Fock
eigenbasis, \cite{michelsen2010-njd}
\begin{align}
  \label{eq:bdge}
  H
  =
  \sum_{n=1}^N \epsilon_n(\varphi) (d_n^\dagger d_n-\frac{1}{2})
  +
  \hbar
  \frac{\dd{\varphi}}{\dd{t}}
  \sum_{m,n=-N}^N M_{mn}(\varphi) d_{m}^\dagger d_{n}
  \,,
\end{align}
where $M_{mn}(\varphi) = -\frac{i}{2} \bra{m,\varphi}[\partial_\varphi
- i\frac{\hat{\tau}_3\partial_\varphi\phi}{2}]\ket{n,\varphi}$
are the connections between the instantaneous
single-particle eigenstates. $\hat{\tau}_3$ is the third Pauli matrix (charge
density operator) in the electron-hole space, and we choose a basis
such that $d_{-n}=d_n^\dagger$, $\epsilon_{-n}=-\epsilon_n$.

The low-energy physics is captured by a model Hamiltonian for
the hybridization of the Majorana states,
\cite{kitaev2001-umf}
\begin{align}
  \label{eq:simple-model}
  H 
  =
  iE_J\cos\left(\frac{\varphi}{2}\right)\eta_1\eta_2
  + 
  i \delta_1 \eta_1\eta_3
  +
  i \delta_2 \eta_4 \eta_2
  \,,
\end{align}
where $\eta_j$ are the four Majorana operators in Fig.~\ref{fig:setup}.
The resulting connections are $M_{1,1}=M_{2,2}=0$, and
\begin{align}
  \label{eq:simple-model-connection}
  M_{\pm2,1}(\varphi)
  = 
  \frac{E_J}{8} \frac{\delta_\mp}{\delta_\mp^2 
    + \left(\frac{E_J}{2}\right)^2 \cos^2(\varphi/2)} 
  \sin\left(\frac{\varphi}{2}\right)
  \,,
\end{align}
where $\delta_\pm=\delta_1\pm\delta_2$.  Away from the level crossing
at $\varphi=\pi$, these matrix elements are proportional to Majorana
state overlaps, which are exponentially small when the length of the
wire segments $L/2$ is large compared to the superconducting coherence
length $\xi=\hbar v_F/\Delta$.

\begin{figure}
  \includegraphics{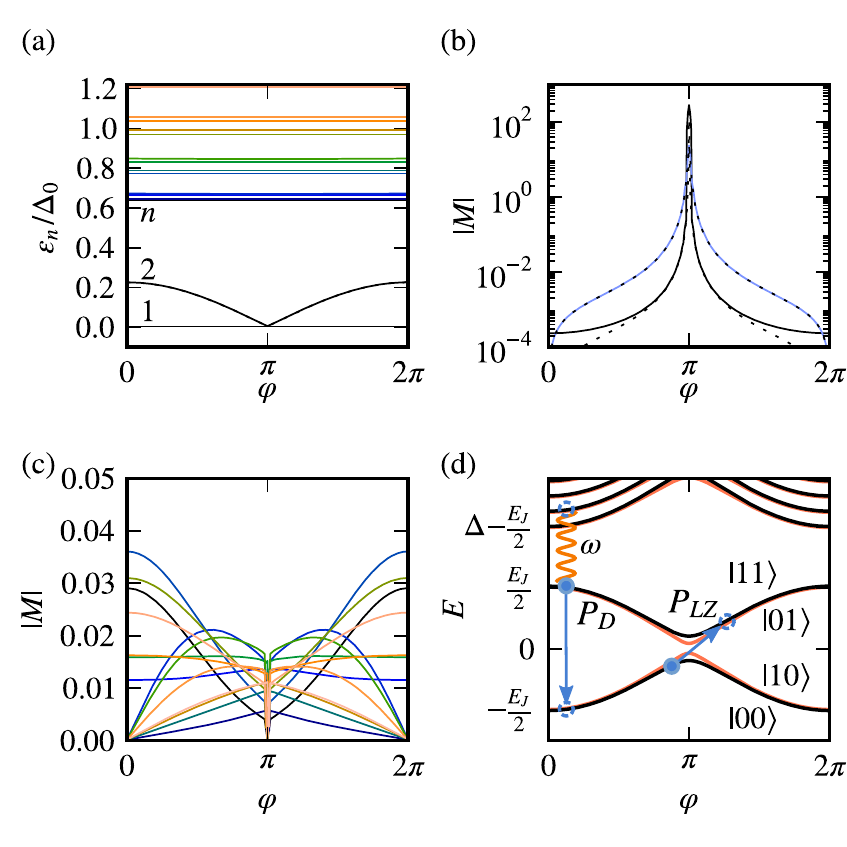}
  \caption{
    \label{fig:spectra}
    (Color online)
    Spectra and connections computed in a Rashba nanowire model.
    (a)
    Andreev spectrum ($\epsilon_n>0$ shown for $n\le18$).
    (b)
    Connections $M_{1,2}$ (black line) and $M_{1,-2}$ (light line)
    as a function of phase difference. Model connections from
    Eq.~\eqref{eq:simple-model-connection} (dotted lines) agree
    closely with the numerical results.
    (c)
    Connections $M_{2,n}$, $2<n\le{}18$ between a low-energy Andreev level
    and the continuum. The other low-energy level is weakly coupled
    to the continuum, $|M_{1,n}|\ll|M_{2,n}|$.
    (d) 
    Schematic many-body spectrum
    with Landau-Zener ($P_{LZ}$) and continuum transitions ($P_D$)
    indicated. Here, 
    $\ket{n_1n_2}$ denote the occupation numbers of levels 1 and 2.
  }
\end{figure}

To compare with a specific microscopic theory, we consider the
connections in a semiconductor nanowire model;
\cite{oreg2010-hla,lutchyn2010-mfa,sau2010-gnp,san-jose2012-aje} the
results in Figs.~\ref{fig:spectra}(a),(b) show how
Eqs.~\eqref{eq:simple-model}, \eqref{eq:simple-model-connection}
capture the low-energy features.  \footnote{ For nanowire Hamiltonian
  $H_{BdG}=[-\frac{\hbar^2}{2m_*}\partial_x^2-\mu +
  i\alpha\sigma_y\partial_x + V_x\sigma_x]\tau_z +
  \Delta_0\sigma_y\tau_+ + \Delta_0^*\sigma_y\tau_-$ in basis
  $(\psi_\uparrow,\psi_\downarrow,\psi_\uparrow^\dagger,\psi_\downarrow^\dagger)$
  we take parameters $E_{so}=\alpha^2m_*/(2\hbar^2)=\unit[50]{\mu e
    V}$, $l_{so}=\hbar^2/(\alpha m_*)=\unit[200]{nm}$, $L=80l_{so}$,
  $\Delta_0=1.7E_{so}$, $\mu=1E_{so}$, $V_x=g\mu_B B_x=10E_{so}$.  The
  junction barrier is implemented by $\mu>\sqrt{V_x^2 -
    |\Delta_0|^2}$.  }  The connections to the continuum at
$\epsilon>\Delta$ are not exponentially small as shown in
Fig.~\ref{fig:spectra}(c), and have a relatively weaker dependency on
$\varphi$.  The energy gap $\Delta-E_J$ depends on the transparency of
the junction. \cite{kwon2004-faj}

Electromagnetic drive couples to the junction by inducing a
voltage $V(t)=-(s_{d}\hbar\omega_d/e)\sin(\omega_dt)$
across the system, equivalent to a
superconducting phase difference
$\varphi(t)=\frac{2e}{\hbar}\int^t\dd{t}V(t)=2s_{d}\cos(\omega_dt)$.

We concentrate on the population dynamics of the current-carrying
low-energy levels, and treat the coupling to the continuum as a perturbation.
\cite{san-jose2012-aje} For simplicity, we assume the continuum
connections $M$ have roughly constant order of magnitude in the
relevant energy range around $\pm\epsilon+\hbar\omega_d$
[cf. Fig.~\ref{fig:spectra}(c)], that the continuum density of states
is steplike, ${\cal N}(\epsilon) = {\cal
  N}\theta(|\epsilon|-|\Delta|)$, and that the quasiparticle
population in the continuum is negligible. The resulting master
equation for the states 1 and 2 is
\begin{align}
  \label{eq:master-equation}
  \dot\rho
  &=
  {\cal L}\{\rho\}
  =
  -i[H_0', \rho]
  -
  \sum_{kk'=-2}^2[A(\epsilon_{k'})+A(\epsilon_{k})^*]
  \\\notag
  &\qquad\times
  \{
  d_{k'} P \rho P d^\dagger_k - \frac{1}{2}[d_k^\dagger d_{k'}, \rho]_+
  \}
  \,,
  \\
  \label{eq:energy-renormalization}
  H_0' 
  &= H_0 
  +
  \frac{i}{2}
  \sum_{kk'=1,2}[A(\epsilon_{k'})-A(\epsilon_{k})^*]
  d_k^\dagger d_{k'}
  \,,
  \\
  \label{eq:continuum-coupling}
  A(\epsilon)
  &\approx
  i
  \hbar
  \Gamma_D
  \log\left(
    \frac{
      (\hbar\omega_d)^2 - (\epsilon - \Delta + i\frac{\hbar\Gamma_0}{2})^2
    }{
      (\hbar\omega_d)^2 - (\epsilon - E_c + i\frac{\hbar\Gamma_0}{2})^2
    }
  \right)
  \,,
\end{align}
where $\Gamma_D=\hbar{\cal N}|M|^2s_d^2\omega_d^2$ is the transition
rate, $P=(-1)^{d_1^\dagger d_1 + d_2^\dagger d_2}$ is the Fermion
parity, and $H_0$ the part of Eq.~\eqref{eq:bdge} involving only
levels $1$ and $2$.  $\Re
A(\epsilon)\approx\pi\hbar\Gamma_D\theta(|\omega_d| -
|\epsilon-\Delta|)$. $E_c$ is a cutoff energy, originating from the
fact that $M\sim{}\epsilon^{-1}$ decays at energies
$\epsilon\gg\Delta$; the results below are insensitive to it.
$\Gamma_0$ is the inverse lifetime of the continuum levels, which are
assumed to be better coupled to external leads than the localized
low-energy bound states. For
$\Delta+\epsilon_2>\hbar\omega_d>\Delta-\epsilon_2$ [see
Fig.~\ref{fig:spectra}(d)], the result describes quasiparticles on
level 2 absorbing energy from the field and escaping to the continuum,
leading to depopulation (``cooling''). \cite{bergeret2010-tom} The
opposite emission process is limited by the low quasiparticle
population in the continuum.

\begin{figure}
  \includegraphics{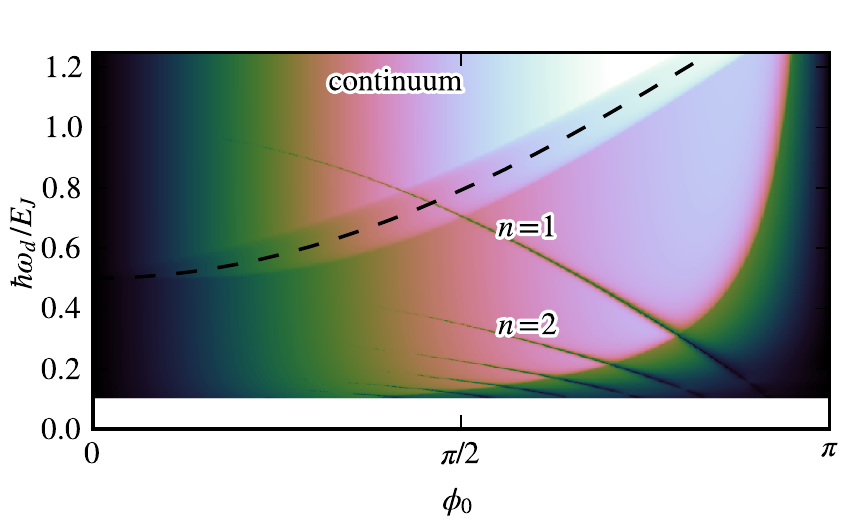}
  \caption{\label{fig:mw} (Color online) Time-averaged current-phase
    relation $\overline{I}(\phi_0)$ as a function of excitation frequency $\omega_d$,
    keeping AC amplitude fixed at $\hbar{}s_d\omega_d=0.1E_J$ and
    $\Gamma_D=10^{-5}E_J$ (lighter color:
    larger current). We take $\Delta/E_J=1.5$ 
    and $\delta_+/E_J=10^{-2}$, $\delta_-/E_J=10^{-3}$.
    The dashed line indicates the threshold $\hbar\omega_d=\Delta -
    \epsilon_2(\varphi)$ for continuum transitions. The $n$-photon
    resonances $n\hbar\omega_d=\epsilon_2(\varphi)-\epsilon_1(\varphi)$
    between the Andreev bound states for $n=1,2$ are indicated.  The
    frequency-dependent threshold near $\varphi=\pi$ is due to LZ
    transitions, which play a role when $\phi_0+2s_d>\pi$.  
  }
\end{figure}

\emph{Current-phase relation.}  Consider first the phase biasing
condition, $\varphi(t)=\phi_0+2s_d\cos(\omega_d t)$, in which the DC
part $\phi_0$ is kept fixed. In this setup, $4\pi$ periodic
nonequilibrium effects are not visible, but one can study
spectroscopic features of the junction.

We augment Eq.~\eqref{eq:master-equation} with quasiparticle poisoning
(parity non-conserving), relaxation (parity-conserving), and dephasing
described by phenomenological rates, for which we assume values
$\Gamma_q, \Gamma_r\sim{}10^{-4}E_J/\hbar$ and
$\Gamma_d\sim{}10^{-3}E_J/\hbar$, respectively.
\cite{supplementaryepaps,san-jose2012-aje,pikulin2012-pad} Resulting
DC current $\overline{\avg{\partial_\varphi H}}$ is shown in
Fig.~\ref{fig:mw}. As in quantum point contacts
\cite{bergeret2010-tom}, resonant transitions can here be identified
with sharp dips in the current-phase relation at $n$-photon resonances
$n\hbar\omega_d=\epsilon_2(\varphi)-\epsilon_1(\varphi)$. The
transition rate is proportional to the connections $M\sim{}\delta/E_J$
(forbidden \cite{lutchyn2010-mfa} if there is no overlap,
$\delta_{1/2}=0$), but it is balanced against the small rates
$\Gamma_q$ and $\Gamma_r$. The second apparent feature is that at
$\Delta+\epsilon_2(\varphi)>\hbar\omega_d>\Delta-\epsilon_2(\varphi)$,
the transitions to continuum \cite{bergeret2010-tom} depopulate level
2 and thereby increase the current.  Note that in conventional quantum
point contacts, the continuum excitation gap $\Delta-\epsilon_2$ is
zero at $\varphi=0$, whereas here it remains finite for all $\varphi$,
reflecting the qualitatively different energy spectrum.

\begin{figure}
  \includegraphics{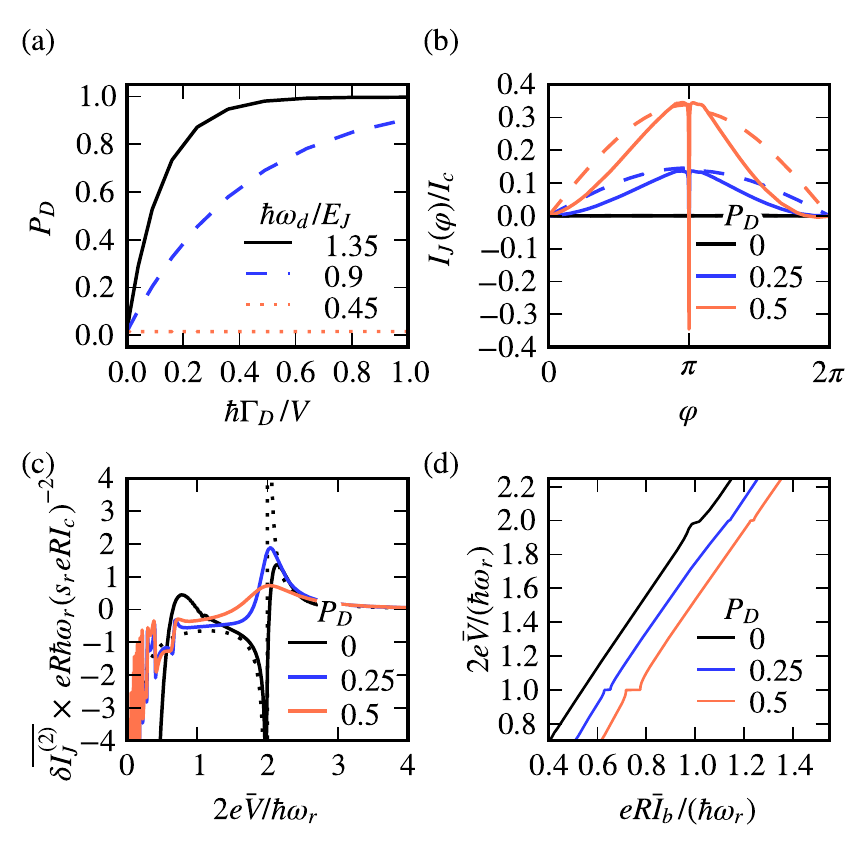}
  \caption{
    \label{fig:shapiro}
    (Color online) 
    Tuning the $4\pi$ periodic Josephson effect.
    (a) Relationship between the effective parameter $P_D$ 
    and the microscopic model, for $\Delta=1.5E_J$
    and $\bar{V}\gg\Gamma_{q,r}$, $\delta^2/E_J$.
    (b) Effective current-phase relation, as obtained from 
    master equation Eq.~\eqref{eq:master-equation}
    with $\hbar\omega_d/E_J=1.35$ (solid lines), and the analytical
    result Eq.~\eqref{eq:minimal-model-cpr} with $P_D$ from Fig.~\ref{fig:shapiro}(a)
    (dashed lines).
    (c) The $4\pi$ Shapiro kink for $P_{LZ}=1$ and $P_D=0,\,0.25,\,0.5$
    from Eq.~\eqref{eq:saddle-point}.
    For $P_D=0$ the result is shown as a dotted line; the solid line indicates
    an exact result 
    without the mean-field approximation.
    (d)
    Composite result from Fig.~\ref{fig:shapiro}(c) and
    Eq.~\eqref{eq:minimal-model-cpr} for the Shapiro steps with parameters
    $RI_c/(\hbar\omega_r)=0.2$, $s_r\equiv{}RI_r/(\hbar\omega_r)=0.5$, 
    $P_{LZ}=1$ and different $P_D$.
    The curves are offset horizontally for clarity.
  }
\end{figure}

\emph{Tuning the $4\pi$ periodic Josephson effect.}  The $4\pi$
periodic Josephson effect \cite{kitaev2001-umf,kwon2004-faj} under dc
bias $\varphi(t)=2e\bar{V}t/\hbar$ in this system requires (i)
Landau-Zener (LZ) transitions at $\varphi(t)=\pi+2\pi n$ and (ii) a
large gap between the continuum and bound state spectra [see
Fig.~\ref{fig:spectra}(d)], which allows correlations between
$\varphi$ and $\varphi+2\pi$ be preserved.  The former requires a
transition probability
$P_{LZ}\simeq{}e^{-8\pi\delta^2/(E_J\bar{V})}\approx{}1$.
\cite{dominguez2012-ddm} The latter can be modified through the
``cooling'' effect discussed above. An additional high-frequency
signal $\Delta+E_J>\hbar\omega_d>\Delta-E_J$ causes a transition to the
lower-energy states with probability $P_D>0$ during a cycle from
$\varphi=2\pi n-\pi$ to $\varphi=2\pi n+\pi$.

The probability $P_D$ can be found
by considering the time evolution of Eq.~\eqref{eq:master-equation} in
the voltage-biased case.  Given initial condition
$\rho(t_1)=\ket{11}\bra{11}$ (or $\ket{01}\bra{01}$) at
$\varphi(t_1)=-\pi+\beta$, at $\varphi(t_2)=\pi-\beta$ we have
$P_D=1-\rho_{01,01}(t_2)-\rho_{11,11}(t_2)$, which is shown in
Fig.~\ref{fig:shapiro}(a). Here, $\beta\gtrsim\delta/E_J$ excludes the
LZ transition. The probability behaves as
$P_D\approx{}1-e^{-c\hbar\Gamma_D/eV}$, $c\sim\pi$ as the transition
rate increases, in agreement with a Landau-Zener type argument, and is
finite for $\hbar\omega_d\gtrsim\Delta-E_J$.

In the experimentally relevant \cite{rokhinson2012-ofa} Shapiro step
experiment, \cite{tinkham1996-its} the system is fed an additional ac
current $I_r\sin(\omega_r t)$, and the $4\pi$ periodicity
manifests as a kink in the dc-current-dc-voltage relation
$\bar{I}(\bar{V})$ at $e\bar{V}=\hbar\omega_r$, whereas the first
$2\pi$ step resides at $e\bar{V}=\hbar\omega_r/2$. Here, we assume the
bias current $I_b(t) = \bar{I} + I_{r}\sin(\omega_r t) + I_d
\sin(\omega_d t)$ contains a further high-frequency component
$\omega_d\gg{}\omega_r$ which induces transitions to the continuum.
To find the Shapiro steps we derive a semiclassical equation
for the phase \cite{tinkham1996-its} by expanding the effective
Keldysh action of the electromagnetic circuit in Fig.~\ref{fig:setup}
to second order in quantum fluctuations $\varphi^q$,
\cite{eckern1984-qds,michelsen2010-njd,sau2012-pfa}
\begin{gather}
  \label{eq:action}
  S[\varphi]
  \simeq-\frac{1}{e}\int_{-\infty}^\infty\dd{t}[\frac{\hbar}{2eR} \dot\varphi^{cl}(t) - I_b(t) + I_J[\varphi^{cl}](t)]\varphi^q(t) 
  \\\notag
  +
  \frac{i}{e^2}\int_{-\infty}^\infty\dd{t}\dd{t'}[\frac{k_BT\delta(t-t')}{R} +
  \frac{P_J[\varphi^{cl}](t,t')}{2}]\varphi^q(t)\varphi^q(t')
  \,.
\end{gather}
The current
$I_J(t)=\avg{\hat{I}(t)}$ and noise
$P_J(t,t')=\frac{1}{2}\avg{\{\hat{I}(t),\hat{I}(t')\}}$ functionals
can in principle be obtained from the master equation.
Considering the
part $\frac{i}{e}\int_{-\infty}^\infty\dd{t'}\theta(t-t')P_J[\varphi^{cl}](t,t')\varphi^q(t')$ in
the last term as a mean field (well-defined on time scales longer than
the relaxation/quasiparticle poisoning time), and assuming the effect
of the junction on the phase dynamics is small ($I_J<\bar{I}_b$,
$P_J<\bar{I}_b^2$), the saddle point becomes
(c.f. Ref.~\onlinecite{sau2012-pfa})
\begin{align}
  \label{eq:saddle-point}
  &
  \frac{1}{2R}
  \partial_t
  \varphi_*(t) + I_J[\varphi_*](t)
  +
  I_J^{(2)}[\varphi_*](t)
  =
  I_b(t)
  \,,
  \\
  \label{eq:mean-field}
  &
  I_J^{(2)}[\varphi_*](t)
  \simeq
  \frac{-i}{e}
  \int_{-\infty}^t\dd{t'}\dd{t''}
  \frac{\delta P_J[\varphi_*](t,t')}{\delta \varphi(t'')}
  D^R(t'',t')
  \,.
\end{align}
The correlation function of the Ohmic environment circuit is
$D^R(t,t')=-\frac{2ie^2R}{\hbar}\theta(t-t')$, $D^R(t,t)=0$. The $4\pi$
periodic memory effects arise solely from the correlation function
$P_J$, as the expectation value $I_J$ must be invariant under
translation $\varphi\mapsto\varphi+2\pi$.

To find analytical results, we evaluate Eq.~\eqref{eq:saddle-point} in
a simplified model retaining only the main physics
(cf. Ref.~\onlinecite{pikulin2012-pad}): we neglect quantum coherence,
and consider only the average populations
$p_+=\frac{\rho_{11,11}+\rho_{01,01}}{2}$,
$p_-=\frac{\rho_{10,10}+\rho_{00,00}}{2}$ of the two upper and lower
many-body states [see Fig.~\ref{fig:spectra}(d)]. The time evolution
is constructed from LZ transitions $p_\pm\mapsto{}(1-P_{LZ})p_\pm +
P_{LZ}p_\mp$ at $\varphi=\pi+2\pi{}n$, and continuum relaxation
$p_+\mapsto{}(1-P_D)p_+$, $p_-\mapsto{}p_-+P_Dp_+$ at
$\varphi=2\pi{}n$.  Under this approximation, the $I_J$ and $P_J$
functionals can be found, \cite{supplementaryepaps} and the IV curve
is obtained from the time average of Eq.~\eqref{eq:saddle-point}.

First, continuum transitions increase the magnitude of the $2\pi$
Shapiro steps, as they remove quasiparticles from the low-energy
levels. This is determined by the effective current--phase relation
$I_J[\varphi](t)$. For slow or small deviations around the trajectory
$\varphi(t)=2\bar{V}t$, it can be approximated by the time-local relation
$I_J(\varphi(t))=\Tr\rho_0(\bar{V},\varphi(t))\hat{I}(\varphi(t))$ in the
periodic steady state $\rho_0(\bar{V},\varphi)=\rho_0(\bar{V},\varphi+2\pi)$. We
find ($\bar{V}>0$)
\begin{align}
  \label{eq:minimal-model-cpr}
  I_J
  &=
  I_c
  P_D
  \frac{
  (1 - P_{LZ})
  \sin(\frac{\varphi}{2})\sgn(\cos(\frac{\varphi}{2}))
  +
  P_{LZ}
  |\sin(\frac{\varphi}{2})|
  }{
    1 - (1-P_D) (1 - 2 P_{LZ})
  }
  \,,
\end{align}
shown in Fig.~\ref{fig:shapiro}(b).  At low excitation
$eRI_r/(\hbar\omega_r)\ll{}1$ the resulting first Shapiro step at
$2e\bar{V}/(\hbar\omega_r)=1$ is similar to the supercurrent step at
$\bar{V}=0$, but with a smaller effective supercurrent
$I_{c,1}=2eRI_JI_r/(\hbar\omega_r)$.  
\footnote{
  Note that we assume a parameter regime outside the enhanced
  $4\pi$ locking effect discussed in Ref.~\onlinecite{dominguez2012-ddm}.  
  However, we expect the effect discussed here has a strong effect
  also in the locked regime.
}

The $4\pi$ periodic features are contained in the time average
$\overline{I^{(2)}_J} = \overline{I^{(2)}_{J0}} + \overline{\delta
  I^{(2)}_{J}}$.  Here,
$\overline{I^{(2)}_{J0}}\simeq{}RI_c^2/(2\bar{V})$, and the part
$\overline{\delta I^{(2)}_J}$ proportional to $I_r^2$ is illustrated
in Fig.~\ref{fig:shapiro}(c). Increasing $P_D>0$ cuts off the
$e\bar{V}=\hbar\omega_r$ resonance in the IV curve.  

For $P_{LZ}=1$ and $P_D\to0$ the correlation function factorizes,
$P_J[\varphi](t,t')=I_c^2\sin(\varphi(t)/2)\sin(\varphi(t')/2)$, so
that Eq.~\eqref{eq:action} can be transformed
\cite{eckern1984-qds,kamenev2010-kta} to the stochastic equation
$\frac{\hbar}{2eR}\partial_t\varphi(t) +
{\cal I}\sin(\varphi(t)/2) = I_b(t)+\xi(t)$ where ${\cal I}$ and the thermal
noise $\xi(t)$ are Gaussian random variables. This result is also shown in
Fig.~\ref{fig:shapiro} for comparison.  Note that phase diffusion due
to a finite temperature also suppresses the Shapiro steps,
which is not taken into account in the figures.

Combining the steps from $I_J$ and $I_J^{(2)}$, we obtain
Fig.~\ref{fig:shapiro}(d), which shows how the IV curve reflects the
change in periodicity from $4\pi$ to $2\pi$ when $P_D$ increases.
This switching is tunable, and can be used to establish both the size
of the gap $\Delta-E_J$ and the way the effect is directly related to
the nonequilibrium state. Namely, in junctions with only $2\pi$
periodic Josephson effect, the high-frequency drive has no effect on
the Shapiro steps within our model.

\emph{Summary.}
In summary, we consider microwave spectroscopy of Josephson junctions
formed from hybridized Majorana bound states.  We discuss what
spectroscopic features manifest in a phase-biased situation, and how
ac excitation can be used to tune the nonequilibrium state in the
$4\pi$ periodic Josephson effect.
High-frequency probing
($\hbar\omega_d\sim{}\unit[400]{mK}\times{}k_B$) was achieved
experimentally in Ref.~\onlinecite{chauvin2007-cfj} for
superconducting quantum point contacts, and we expect high-frequency
manipulation of junctions with Majorana bound states would also be
feasible.

This work is supported by the Academy of Finland, DFG grant RE
2978/1-1 and the EU FP7 project SE2ND.

\appendix

\begin{widetext}

\section{Phenomenological rates}  

Quasiparticle poisoning ($q$), relaxation ($r$) and dephasing ($d$)
can be described with the master equation Eq.~(4) of the main text, by
adding phenomenological rate terms:
\cite{pikulin2012-pad,san-jose2012-aje}
\begin{align}
  {\cal L}_q\{\rho\}
  &=
  \Gamma_q \sum_{k=-2}^2 {\cal X}(d_kP)\{\rho\}
  \,,
  \\
  {\cal L}_r\{\rho\}
  &=
  \Gamma_r [{\cal X}(d_2d_1) + {\cal X}(d_1^\dagger d_2)]\{\rho\}
  \,,
  \\
  {\cal L}_d\{\rho\}
  &=
  \Gamma_d \sin^2\bigl(\frac{\varphi}{2}\bigr)
  \sum_{k=1,2}  {\cal X}(d_k^\dagger d_k)\{\rho\} 
  \,,
  \\
  {\cal X}(A)\{\rho\}
  &\equiv 
  A \rho A^\dagger - \frac{1}{2}[A^\dagger A, \rho]_+
  \,.
\end{align}
Similar terms can also be derived by integrating out a bosonic
bath. \cite{cheng2012-tpm} The dephasing rate is proportional to
$|\partial_\varphi\epsilon_2(\varphi)|^2$, hence the phase dependence.
Assuming QP poisoning time of order $\unit[1]{\mu s}$ and a Josephson
energy of $E_J=\unit[100]{mK}$, we have
$\hbar\Gamma_q\simeq{}10^{-4}E_J$. With environmental impedance
$R\sim\unit[1]{\Omega}$, the relaxation rate $\Gamma_r$ is also
expected to be of a similar magnitude. \cite{san-jose2012-aje} In the
main text, we assumed the dephasing rate is
$\hbar\Gamma_d=10^{-3}E_J$.  However, as long as
$\hbar\Gamma_d\ll{}\hbar\omega_d,E_J$, dephasing affects the results
here only slightly because they do not require quantum coherence.

\section{Simplified master equation}

Below, we discuss a simplified model, which describes the
quasiparticle physics of the Andreev bound states on a semiclassical
level. This enables us to obtain analytical results for the Shapiro
steps, and we can compare its predictions to those of the more
detailed microscopic model.

First, we assume the dephasing in the system is large enough, so that
quantum coherence plays no role on time scales of $\hbar/(e\bar{V})$
(i.e. between two consequent Landau-Zener transitions). Under this
assumption, the density matrix is projected as
$\rho\mapsto\mathop{\mathrm{diag}}\rho$.  Further, assuming that the
lower and upper two many-body levels are equivalent (i.e.,
$\epsilon_1\approx0$) the system is described by the populations
$p_+=(\rho_{11,11}+\rho_{01,01})/2$ and
$p_-=(\rho_{10,10}+\rho_{00,00})/2$. This defines a linear projection
superoperator $P\{\rho\} = (p_+, \; p_-)^T$.  Effect of Landau-Zener
(LZ) and continuum (D) transitions can then be taken into account with
projected propagators $U(t,t')=P{\cal T}e^{\int_{t'}^t\dd{t''}{\cal
    L}(t'')}P^T$ of the master equation. For time evolution between
$\varphi(t')=\pi-\beta\mapsto{}\varphi(t)=\pi+\beta$ and
$\varphi(t)\mapsto{}\varphi(t'')=3\pi-\beta$ with
$\beta\sim\delta/E_J\ll2\pi$ we have:
\begin{align}
  U_{LZ}
  \equiv
  P U(t,t') P^T
  \simeq
  \begin{pmatrix} 1 - P_{LZ} & P_{LZ} \\ P_{LZ} & 1 - P_{LZ}\end{pmatrix}
  \,,
  \quad
  U_{D}
  \equiv
  P U(t'',t) P^T
  \simeq
  \begin{pmatrix} 1 - P_{D} & 0 \\ P_{D} & 1\end{pmatrix}
  \,.
\end{align}
As a further simplification, we assume below that the continuum
relaxation occurs instantaneously when the phase difference crosses
the points $\varphi=2\pi n$.  The current superoperator
$\hat{I}\{\rho\}=P\frac{e}{2\hbar}[\partial_\varphi{}H,\rho]_+P^T$ in this
representation is
\begin{align}
  \hat{I} \simeq 
  \frac{e}{\hbar}
  \begin{pmatrix}
    \partial_\varphi\epsilon_2(\varphi) & 0 \\
    0 & -\partial_\varphi\epsilon_2(\varphi)
  \end{pmatrix}
  \,.
\end{align}
This allows straightforward approximation of the correlation functions
required for the effective action.

Of importance here is the time evolution over a single period,
$\varphi\mapsto\varphi+2\pi$ (assuming $\varphi(t)$ is monotonically
increasing):
\begin{align}
  U_c(\varphi)
  =
  U(\varphi+2\pi,\varphi)
  =
  \begin{cases}
    U_D U_{LZ}
    \,,
    &
    0<\varphi<\pi
    \\
    U_{LZ} U_D
    \,,
    &
    \pi<\varphi<2\pi
  \end{cases}
  \,.
\end{align}
This determines the cyclic steady state $v_0=P\{\rho_0\}$ of the system:
$U_c(\varphi)v_0(\varphi)=v_0(\varphi)$, and the transient state
$U_c(\varphi)v_1(\varphi)=\lambda_1v_1(\varphi)$.  Here, the
decay factor for the transient state is $\lambda_1 =
(1-P_D)(1-2P_{LZ})$ and is independent of $\varphi$.  Computing
$I_J=\Tr\hat{I}\rho_0 = 1^T \hat{I} v_0$ we find Eq.~(10) of
the main text.

It is now convenient to define the projection superoperators $Q_0(\varphi)
= v_0(\varphi) 1^T$ and $Q_1(\varphi) = 1 - Q_0(\varphi)$ to the
steady-state and transient subspaces. These satisfy $Q_0(\varphi)
U_c(\varphi) = U_c(\varphi)Q_0(\varphi) = Q_0(\varphi)$, $Q_1(\varphi)
U_c(\varphi) = U_c(\varphi)Q_1(\varphi) = \lambda_1 Q_1(\varphi)$,
$Q_0(\varphi)^2 = Q_0(\varphi)$, and $Q_1(\varphi)^2 = Q_1(\varphi)$.

We assume the drive
\begin{align}
  \label{eq:minimal-model-drive}
  \varphi_*(t) = \phi_0 + 2e\bar{V}t/\hbar + 2s_r[\cos(\omega_r t + \phi_r) - \cos(\phi_r)]
  \,.
\end{align}
When $e\bar{V}$ and $\hbar\omega_r$ are not exactly commensurate,
long-time averages of periodic functions of $\varphi_*(t)$ can be
computed by averaging over $\phi_0$ and $\phi_r$.  The frequency
component $\omega_d$ inducing the continuum transitions does not
directly couple to the low-frequency dynamics, and is not included
here for simplicity.

We can now evaluate the time-average of Eq.~(9) of the main text under
drive~\eqref{eq:minimal-model-drive}. First, we remark that
$\frac{\delta}{\delta \varphi^{cl}(t'')} P U(t,t') P^T \approx 0$ within our
approximations (see also Ref.~\onlinecite{sau2012-pfa}).  Moreover, $D^R(t,t)=0$, so that only one term
contributes in the expression
\begin{align}
  D^R(t'',t')
  \frac{\delta}{\delta \varphi^{cl}(t'')} 
  \avg{\hat{I}(t)\hat{I}(t')}
  =
  D^R(t'',t')
  \avg{(\partial_\varphi \hat{I})(t)\hat{I}(t')}
  \delta(t-t'')
  \,.
\end{align}
Evaluating now the long-time average and changing integration
variables, we find
\begin{align}
  \label{eq:minimal-model-voltage-shift}
  \overline{
    I_{J}^{(2)}[\varphi_*](t)
  }
  &=
  -2R
  \int_{0}^{2\pi}\frac{\dd{\phi_0}}{2\pi}
  \int_0^{2\pi}\dd{\theta}
  \Tr[
  (\partial_\phi\hat{I})(\theta+\phi_0)
  U(\theta+\phi_0, \phi_0)
  Q_1(\phi_0)
  \hat{I}(\phi_0)
  Q_0(\phi_0)
  ]
  G(\theta)
  \,,
  \\
  \notag
  G(\theta)
  &=
  \frac{e}{\hbar}
  \sum_{m=0}^\infty
  \int_{0}^{2\pi}\frac{\dd{\phi_r}}{2\pi}
  \frac{
    (1 - P_D)^m
    (1 - 2P_{LZ})^m
  }{
    \partial_t\varphi_*(t_*(\theta+2\pi m, 0, \phi_r))
  }
  \\
  &=
  \frac{1}{2V}
  \frac{1}{1 - (1-P_D)(1 - 2P_{LZ})}
  +
  \frac{(s_r\hbar\omega_r)^2}{4\bar{V}^3}
  \frac{
    \cos(\frac{\theta \hbar\omega_r}{2e\bar{V}})
    -
    (1-P_D)(1 - 2P_{LZ})\cos(\frac{(2\pi-\theta)\hbar\omega_r}{2e\bar{V}})
  }{
    \sin(\frac{\pi\hbar\omega_r}{e\bar{V}})^2
    +
    [
    \cos(\frac{\pi\hbar\omega_r}{e\bar{V}})
    -
    (1-P_D)
    (1 - 2P_{LZ})
    ]^2
  }
  +
  {\cal O}(s_r^4)
  \,.
\end{align}
The function $t_*(\theta,\phi_0,\phi_r)$ is defined by
$\varphi_*(t_*(\theta,\phi_0,\phi_r))=\theta$.

All information about the $4\pi$ periodicity and the drive is
contained in the factor $G(\theta)$. For $P_{LZ}=0$, the maximum
amplitude is obtained when the drive is $2\pi$ periodic in $\theta$,
and for $P_{LZ}=1$ when it is $4\pi$ periodic. Moreover, $P_D>0$
suppresses correlations between cycles, and cuts them off completely
at $P_D=1$.

Equation~\eqref{eq:minimal-model-voltage-shift} can be evaluated
analytically, in the leading order in the drive amplitude $s_r$.
In the interesting case of $P_{LZ}=1$, we obtain
\begin{align}
  \overline{I_J^{(2)}} 
  &= \overline{I_{J,0}^{(2)}} + \overline{\delta I_{J}^{(2)}} + {\cal O}(s_r^4)
  \\
  \overline{I_{J,0}^{(2)}}
  &=
  \frac{
    2 (1 - P_D) [(4 - \pi) P_D + 2 \pi]
  }{
    (2 - P_D)^3 \pi
  }
  \frac{RI_c^2}{\bar{V}}
  \\
  (eR)^{-1} I_c^{-2} \overline{\delta I_J^{(2)}}
  &=
  -\frac{2 (1-P_D) P_D^2 s_r^2 (\hbar\omega_r)^3 \sin \left(\frac{\pi\hbar\omega_r
   }{e\bar{V}}\right)}{\pi  (P_D-2)^2 (e\bar{V}-\hbar\omega_r )^2 (e\bar{V}+\hbar\omega_r
   )^2 \left(P_D^2-2 P_D \cos \left(\frac{\pi\hbar\omega_r
   }{e\bar{V}}\right)-2 P_D+2 \cos \left(\frac{\pi\hbar\omega_r
   }{e\bar{V}}\right)+2\right)}
  \\\notag&\quad
  -\frac{4 (1-P_D) P_D s_r^2 (\hbar\omega_r)^2 \cos
   \left(\frac{\pi\hbar\omega_r }{2e\bar{V}}\right)}{\pi  (P_D-2)e\bar{V}
   (e\bar{V}-\hbar\omega_r ) (e\bar{V}+\hbar\omega_r ) \left(P_D^2-2 P_D \cos
   \left(\frac{\pi\hbar\omega_r }{e\bar{V}}\right)-2 P_D+2 \cos \left(\frac{\pi 
   \hbar\omega_r }{e\bar{V}}\right)+2\right)}
  \\\notag&\quad
  + \frac{(1-P_D) s_r^2 (\hbar\omega_r)
   ^2}{(P_D-2)^2 e\bar{V} (e\bar{V}-\hbar\omega_r ) (e\bar{V}+\hbar\omega_r )}
  \,.
\end{align}
This result is plotted in Fig.~4 of the main text.  Note that it
applies in the limit where the junction has a small effect on the
dynamics of the phase $\varphi$ --- that is,
Eq.~\eqref{eq:minimal-model-drive} is valid. Divergences
(e.g. $\bar{V}\to0$) indicate a breakdown of this approximation.

\section{Discrete continuum spectrum}

The level spacing in the continuum part of the energy spectrum in
nanowires is not necessarily small, for instance in a proximity
nanowire setup where the effective energy gap $\Delta$ in the nanowire
is smaller than the gap $\Delta_S$ of the proximity superconductor
inducing it.  Our main results apply also in this case ---
Figs.~\ref{fig:supp1} and \ref{fig:supp2} show that only limited
qualitative changes are expected from the discreteness of the
continuum spectrum.

\begin{figure}
  \includegraphics{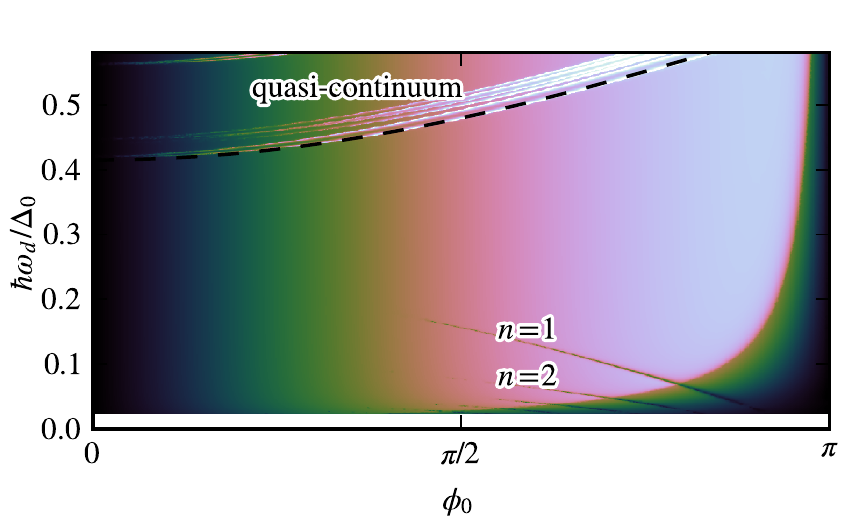}
  \caption{\label{fig:supp1}
    Time-averaged current-phase relation for a nanowire system.
    As Fig.~3 in the main text, but for the nanowire 
    system whose spectrum and connections are displayed in Fig.~2,
    corresponding to parameters given in [28].
    The wire was chosen long, so that the energy splittings
    $\delta_\pm$ of the Majorana states
    are small, and the resonances corresponding to transitions
    between the low-lying bound states are weak.
    They are more pronounced in shorter wires, as illustrated
    in Fig.~2 of the main text.
    Resonances associated with transitions to continuum levels
    are not similarly suppressed.
    The assumed continuum inverse lifetime is $\Gamma_0/E_J=10^{-3}$.
  }
\end{figure}

\begin{figure}
  \includegraphics{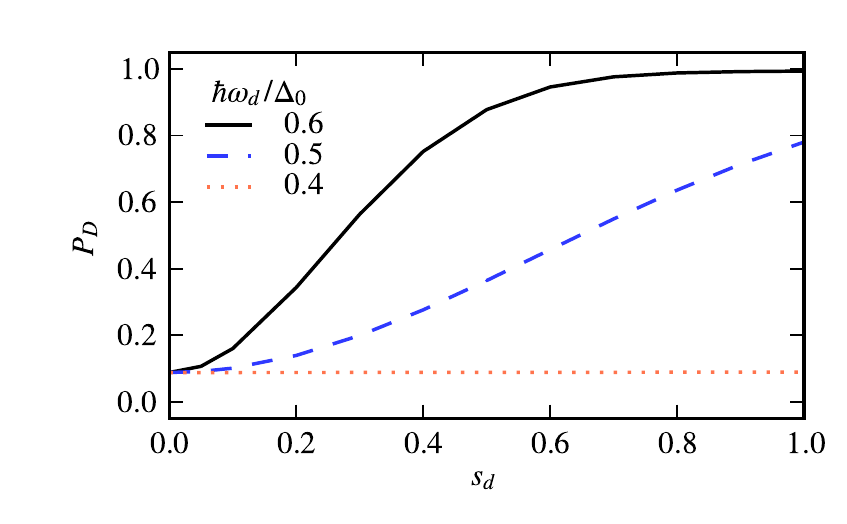}
  \caption{\label{fig:supp2}
    Probability of continuum relaxation for a nanowire system.
    As Fig.~4(a) in the main text, but for the nanowire 
    system whose spectrum and connections displayed in Fig.~2,
    and shown as a function of $s_d$ with fixed $eV=0.01E_J$.
    The offset from zero is due to a finite spontaneous quasiparticle
    poisoning and relaxation with rates $\Gamma_{r,q}/E_J=10^{-4}$.
    The result is not sensitive to continuum lifetime $\Gamma_0^{-1}$.
  }
\end{figure}

\end{widetext}
\end{document}